\def\eeq{\end{equation}}

\def\beq{\begin{equation}}

\def\bea{\begin{eqnarray}}

\def\eea{\end{eqnarray}}

\documentstyle[aps,prl,epsfig]{revtex}

\begin{document}

\title{Cosmology in a non-standard statistical background}

\author{Diego F. Torres${^{1,2}}$
\thanks{dtorres@venus.fisica.unlp.edu.ar} and 
H\'ector Vucetich${^2}$}

\address{${^1}$ Astronomy Centre, CPES, University of Sussex, Falmer,
Brighton BN1 9QJ, United Kingdom\\
${^2}$Departamento de F\'{\i}sica, 
Universidad Nacional de La Plata,
C.C. 67, 1900, La Plata,  Argentina}

\maketitle

\begin{abstract}
We study the primordial neutron to baryon ratio in a cosmological
expanding background within the context of non-extensive statistics,
in a fully analytical approach.
First order corrections to the weak interaction rates
and energy densities in the early universe are obtained 
and their consequences on the Helium synthesis are analized.
We find that, if the nucleosynthesis scenario should be in agreement with observation,
this entails a very restrictive bound upon 
$q$, the non-extensive parameter.
We also study some cosmological
interesting situations such as the moment of pair annihilation,
conserved comoving quantities
and derive for the nonextensive statistics, temperature relationships,
before and after the removal of pairs.

{\it PACS number(s):} 05.30.-d, 05.70.-a, 26.35.+c, 98.80.Ft

\end{abstract}


\section{Introduction}

An interesting generalization of the Boltzmann-Gibbs
entropy form has been recently proposed by Tsallis \cite{t2}, 
for a recent review see \cite{t1,PHYSA}, for a full bibliography see \cite{WEB}.
This new entropy, that we discuss below, posseses the usual
properties of positivity, concavity and irreversibility and
generalize the additivity in a non-extensive way.
In the later years, a lot of work has been done in order to show either
that many relationships involving energy and entropy in the usual
scheme holds in the new one  or  in order 
to search for suitable generalizations for them.
We should mention, among others, 
the work of Curado and Tsallis \cite{t3}, that shows
that the whole mathematical structure of thermodynamics --Legendre
transform based-- becomes invariant under a change from extensive to
non-extensive statistics (NES).  This, in fact, is a property recently proved
to be valid for any entropic functional form \cite{PLAS-ENTRO}, making then
important to have a partition function within the formalism.
Boltzmann-Gibbs thermostatics constitutes very complete and
powerful techniques in situations whenever thermodynamics extensivity
holds. But this formalism provide divergent partition, energy and
entropy functions whenever the physical system includes long range
forces or long memory effects. In this regard, 
the connection with
quantum mechanics and with information theory has  been stablished
\cite{t4} and further applications to self-gravitating objects \cite{t5} and
astrophysics \cite{t6}, to L\'evy flights \cite{t7}, turbulence of fluids \cite{BOGHO}
and to the 
problem of solar neutrinos \cite{solar} have also been studied.
In general, systems which present a fractal like or unconventional
structure in their space-time description or in their phase space evolution
could develop hardly tractable mathematical problems within the standard
formalism \cite{PHYSA}.
Very recently, Grigera \cite{GRIGERA} derived a molecular dynamic test for systems
of long range interacting particles with potentials of the form $A/r^{12}-B/r^{\alpha}$.
He found that the potential energy per particle do not converge 
for some cases of $\alpha$
and sizes of the system, whereas non-extensive quantities do. Also see Refs.
\cite{KIM,SAMPAIO}.

Non-extensive thermostatistics was also used to study
the cosmic blackbody radiation \cite{Tsallis3,P2V}. 
This was the setting in which powerful bounds upon the non-extensive
parameter were obtained. 
We shall review some of these results in our final section.
A new analysis of NES blackbody cosmological radiation showed that,
independently of the degree of non-extensivity, the
temperature of the
universe varies as the inverse of the scale factor of a flat
Friedmann-Robertson-Walker metric \cite{Barraco}. This result
suggested cosmology of the early universe as a very profitable 
arena for searching  observational consequences of NES. 
As far as we know,
the first of such studies were presented in \cite{TORRES_T1}, 
where a stringent bound on non-extensivity, 
of comparable order of magnitude as those obtained via 
the cosmic blackbody radiation, was derived.
There, it was analized how a first order variation in the non-extensive parameter
would affect the amount of primordial Helium production, a quantity which can
be compared with observational results, via a rather easy and straightforward
approximate technique.

We now turn to the issue of early universe cosmology
and particularly, to the computation of primordial neutron  to
baryon ratio in the framework of NES. We shall also analyze 
cosmological situations of great importance, such as pair
annihilation, temperature relationships, and comoving quantities.
We thus refine
the previous study and enhance it by a sharpening of the model.
Although we shall present a very brief review of the main formulae of NES; in 
first approach, this work should be understood as the search of
deviations from the standard model of the early universe
provided by a different statistical description.

What we want to use is a direct analytical computation of Helium production, following
the work of Bernstein, Brown and Feimberg \cite{Bernstein}
for the standard model, to allow for an approximate 
study. It is a matter of fact, that even when nucleosynthesis processes
may be as complicated as one can afford, some simple assumptions
makes results agree, within a typical few percent, with those
generated by a numerical code, thus allowing for the isolation
of the essential physics. That is exactly what it is needed here
in order to study how, for instance, a different formula
for the number of particles influence the evolution of the early
universe and how can this be used to extract testeable
predictions. Nucleosynthesis has given a strong basis to do such a thing
when alternatives theories of gravitation are concerned, see for instance
\cite{Casas2,TORRES_N,TORRES_W}, and also for particle
physics \cite{Steigman,Kolb,Lucchin} 
and we hope to show, the same is applicable
for alternative statistics.

The rest of the work is organized as follows. Firstly, some useful
concepts and formulae of NES are presented.
Afterwards, the analytical computation of the neutron to baryon ratio
is done, and finally we study the necessary approach to handle with
neutron decay corrections. 
In the final section, we can use our previous results to estimate the mass fraction
of primordial Helium produced in the context of NES. 
Although we shall not have at that moment every ingredient needed to do such
a thing with great precision, the bound which we obtain hardly constrain the
degree of  non-extensivity in the very early
universe and is already comparable with previous results.
Finally, we state our conclusions and mention 
other recently obtained related bounds.

\section{Nonextensive statistics}

Let us first recall some formulae of nonextensive statistic (NES), 
in Tsallis's approach. The formalism starts by postulating \cite{t2}

{\bf Postulate 1}.- {\it The entropy of a system that can be found
with probability
$p_i$ in any of  $W$ different microstates $i$ is given by}

\beq
S_q=k\frac {1}{(q-1)}\, \sum_{i=1}^W [p_i\,-\,p_i^q]=
k \frac {1}{q-1} \left( 1-\sum_{i=1}^{W} p_i ^q\right), 
\label{Sq}
\eeq
{\it with  $q$ a real parameter}. We have a different statistics for
every
possible $q-$value. In (\ref{Sq}) we have used, of course, that

\beq
\sum_i p_i =1.
\eeq
In general,

\beq
\label{Sq2}
S_q=k \frac{1-{\rm Tr} \rho^q}{q-1}.
\eeq

{\bf Postulate 2}.- {\it An experimental measurement of an observable
$A$,
whose
expectation value in microstate $i$ is $a_i$, yields the $q$-
expectation
value} (generalized expectation value (GEV))

\beq
\label{gev}
<\,A\,>_q\,=\,\sum_{i=1}^W\,p_i^q \,a_i = {\rm Tr} \rho^q \hat A,
\eeq
{\it for the observable} $A$.

These two statements have the rank of {\it axioms}. As such, their 
validity is to be decided exclusively by the
conclusions to which they lead, and ultimately by comparison with
experiment.
The entropy given in (\ref{Sq}) is non-negative and reproduces the
Boltzmann-Gibbs one ($S_1=k \sum_i p_i \ln p_i$) in the limit $q \rightarrow 1$.
In fact, note that (\ref{Sq}) can be obtained from the Boltzmann-Gibbs expression by writing
$\ln (x)=\lim_{q\rightarrow 1} (x^{q-1} -1)/(q-1)$ and then dropping the limit operator.
It satisfies the pseudoadditivity property which states that if $A$ and $B$
are two independent systems ($p_{ij}^{(A+B)}=p_i^A p_j^B$) then,

\beq
\frac {S_q(A+B)}{k} = \frac{S_q(A)}{k} + \frac {S_q(B)}{k} + (1-q) \frac{S_q(A)}{k}
\frac{S_q(B)}{k},
\eeq
which shows that $(1-q)$ is a measure of the degree of non-extensivity of the 
system. The optimization of $S_q$ given by (\ref{Sq2}) together with the
constraints ${\rm Tr } \rho=1$ and ${\rm Tr} \rho^q \hat H=U_q < \infty$ yield
to the canonical ensemble equilibrium distribution \cite{t2,t1},

\begin{equation}
\hat \rho=
\frac{1}{Z_q}\left[ \hat 1 - (1-q) \beta \hat H\right]^\frac{1}{1-q},
\end{equation}
and to the generalized partition function,

\begin{equation}
Z_q=Tr \left[ \hat 1 - (1-q) \beta \hat H\right]^\frac{1}{1-q}.
\end{equation}
Here, as usual,
$\beta = 1/kT$ and $\hat H$ is the hamiltonian of the system.
We shall focus in the $\beta (q-1)\rightarrow 0$ limit, 
in which a first order expansion
allows analytical computations. The expression of  a general mean value
of an operator was computed in this limit by Tsallis, Sa Barreto and Loh 
\cite{Tsallis3}. When applied to particle number operators, the result is:

\begin{equation}
\label{<n>}
<\hat n>_q= <\hat n>_{BG}  Z_{BG}^{q-1}
\left[ 1+ (1-q) x \left[ \frac{<\hat n^2>_{BG}}{<\hat n>_{BG}} +
\frac x2 \left( <\hat n^2>_{BG} - \frac{<\hat n^3>_{BG}}{<\hat n>_{BG}} \right)
\right] \right],
\end{equation}
where $x$ stands for $\epsilon/kT$ ($\epsilon$ is the energy of a simple
particle) and the symbol $BG$ means {\sl to be computed in Boltzmann-Gibbs
statistics}.
With the standard values of $<\hat n^2>_{BG}$ and $<\hat n^3>_{BG}$, 
for fermions and bosons, we obtain the main 
corrections (the second terms in equation (8)), $C_{bosons,fermions}$, as:

\begin{equation}
\label{correc-bos}
C_{bosons} = \frac{1}{\exp{(x)}-1} x \left( \frac{1+\exp{(-x)} } {1-\exp{(-x)}} - 
\frac{x}{2}\frac{1+3\exp{(-x)} } {(1-\exp{(-x)})^2} \right)
\end{equation}
\begin{equation}
\label{correc-fer}
C_{fermions} = \frac{1}{\exp(x)+1} x \left(1+\frac x2 \left(
\frac{1}{1+\exp{(x)}} - \frac{1+\exp{(2x)}+2\exp{(x)} } {(1+\exp{(x)})^2} 
\right) \right)
\end{equation}
When $x$ is such that $\exp{(x)} \gg 1$, then, $C_{bosons} \simeq
C_{fermions} \simeq x(1-x/2) \exp{(-x)}$; while for $x$ big enough
we get $C_{bosons} \simeq C_{fermions} \simeq (-x^2/2) \exp{(-x)}$.

\section{Neutron-Proton Ratio in an Expanding Background}

We turn now our attention to the description of the model for the evolution
of the neutron abundance as the universe evolves. As stated in the introduction,
we shall follow the leading ideas of Bernstein, Brown and Feimberg 
\cite{Bernstein}.
We shall denote by $\lambda_{pn}(T(t))$ 
the rate for the weak processes to convert
protons into neutrons and by $\lambda_{np}(T(t))$  the rate for the reverse ones.
$X(T(t))$ will be, as usual, the number of neutrons to the total number
of baryons. For it, a kinetic equation may be built,

\begin{equation}
\frac{dX(t)}{dt}= \lambda_{pn}(T) (1-X(t)) - \lambda_{np}(T) X(t).
\end{equation}                                                    
The solution to the previous equation reads:

\begin{equation}
\label{X2}
X(T)= \int_{t_0}^t  dt^\prime I(t,t^\prime) \lambda_{pn}(t^\prime) + X(t_0) I(t,t_0),
\end{equation}                                                    
with the integrating factor given by,

\begin{equation}
I(t,t^\prime) = \exp {\left( - \int_{t^\prime}^t    d\hat t \Lambda(\hat t) \right)}
\end{equation}                                                    
where,

\begin{equation}
\Lambda(t)= \lambda_{pn}(t) + \lambda_{np}(t) .
\end{equation}                                                    
Note that the form of this 
solution does not depend on the statistical mechanics used. However, changing
the statistical mechanics entails modifications in the rates, and this will
make the solution differs from the standard.
Expression (\ref{X2}) is simplified by setting the initial value $t_0=0$.
We expect that $\lambda_{np}$ and $\lambda_{pn}$ be very large at early
times and high temperatures. Then, the integration factor $I(t,t_0)$ will be very
small for times order $t \simeq 1/\Lambda(t_0)$ and therefore the term
$X(t_0) I(t,t_0)$ can be omitted. Besides, we also expect that due to the large
interaction rates, the integral in the first term will be not sensitive to the change
of $t_0$ by 0.
With this choice, we now have,

\begin{equation}
\label{X3}
X(t)= \int_{0}^t dt^{\prime}  I(t,t^\prime) \lambda_{pn} .
\end{equation}                                                    
Finally, we may note that

\begin{equation}
I(t,t^\prime) =\frac{1}{\Lambda(t^{\prime}) } \frac{d}{dt^{\prime}} I(t,t^\prime),
\end{equation}                                                    
and integrate by parts to obtain

\begin{equation}
\label{X4}
X(t)= \frac {\lambda_{pn}(t)}{\Lambda(t)}-
\int_{0}^t  dt^{\prime} I(t,t^\prime) \frac{d}{dt^{\prime}} \left( 
\frac{\lambda_{pn}(t^{\prime})}{ \Lambda(t^{\prime})}\right).
\end{equation}                  

\subsection{Rate Formulae}

To explicitly compute expression (\ref{X4}), we need to know the functional
form for the rates. Let us first consider
the rate $\lambda_{np}(t)$. This is the sum of the rates of three individual
processes,

\begin{equation}
\lambda_{np}=\lambda_{\nu+n \rightarrow p+e^+}+
\lambda_{e^+ +n \rightarrow p+ \hat \nu}+
\lambda_{n \rightarrow p+e^- + \hat \nu}
\end{equation}
which are given by \cite{Weimberg},

\begin{equation}
\label{r1}
\lambda_{\nu+n \rightarrow p+e^-}=A \int_0 ^\infty dp_{\nu} p_{\nu}^2
p_e E_e (1-<\hat n_e>)<\hat n_{\nu}>
\end{equation}
\begin{equation}
\label{r2}
\lambda_{e^+ +n \rightarrow p+\hat \nu}=A \int_0 ^\infty dp_{e} p_{e}^2
p_{\nu} E_{\nu} (1-<\hat n_{\nu}>)<\hat n_{e}>
\end{equation}
\begin{equation}
\label{r3}
\lambda_{n \rightarrow p+e^- + \hat \nu}=A \int_0 ^{p_{0}}dp_{e} p_{e}^2
p_{\nu} E_{\nu} (1-<\hat n_{\nu}>)(1-<\hat n_{e}>),
\end{equation}
where $A$ is an overall effective constant --fixed by the experimental value of
$\lambda_{n \rightarrow p+e^- + \hat \nu}$--, $p_{\nu ,e}$ stands for the magnitudes
of the neutrino and electron momentum and $E_{\nu ,e}$ for their
energies. The recoil energy of the nucleons may be neglected, and this enable us to write the energy
conservation equation $E_\nu  +m_n=E_e +m_p$ for (\ref{r1}) and 
$E_\nu  +m_p=E_e +m_n$ for (\ref{r2}). These equations must be used in order
to explicitly compute the integrals. In (\ref{r3}), $E_\nu = \Delta m-E_e>0$,
--with $\Delta m=m_n-m_p=1.29 MeV$-- and this gives the upper
limit in the integration range. Finally, $n_{\nu ,e}$ are distributions
functions and $(1-n_{\nu ,e})$ are blocking Pauli factors. 
In the usual scheme, $n_{\nu ,e}$ are given by Fermi distributions,
whereas in our case, by the $<\hat n_{\nu,e}>_q=
n_{\nu, e }(q)$ of the previous section.
In general, the electron and neutrino temperatures, $T_e$ and $T_\nu$,
may differ because at the end of the freezing out period, electrons and positrons
annihilate, heating only the photons and mantaining with them thermal
equilibrium. The magnitude of this difference will be discussed below.
As we shall see, this difference will be of a few percent during the period
of consideration and our first approximations, following Bernstein et. al.,
is to set all temperatures equal, $T=T_e=T_\nu=T_\gamma$.
When working in the Boltzmann-Gibbs framework this unique assumption
ensure that the rates for reverse reactions, such as $e^- + p \rightarrow
n+ \nu$, obey the principle of detailed balance (PDB), that we analize below
in this general setting.

\subsection{Principle of Detailed Balance}

Let us consider, in the Boltzmann-Gibbs (BG) statistics and as an example,
the reverse rate for (\ref{r1}), that is,

\begin{equation}
\label{reverse1}
\lambda_{e^- +p \rightarrow n + \nu}=A \int_0 ^\infty dp_{e} p_{e}^2
p_\nu E_\nu (1-<\hat n_\nu>)<\hat n_{e}>.
\end{equation}
Using energy conservation and the fact that $p_e dp_e = E_e dE_e$
we have, neglecting the recoil energy of the nucleon, $dE_e=dE_\nu=
dp_\nu$, and thus:

\begin{equation}
\label{reverse2}
\lambda_{e^- +p \rightarrow n + \nu}=A \int_0 ^\infty dp_{\nu} p_{\nu}^2
p_e E_e (1-<\hat n_\nu>)<\hat n_{e}>.
\end{equation}
As $<\hat n_{\nu ,e}>=n_{\nu ,e}$ are Fermi-Dirac distributions, it is easily seen that 
$(1-n_{\nu})=n_\nu  \exp{(E_\nu /T)}$ and that $n_e=(1-n_e) \exp{(-E_e /T)}$.
Replacing this into (\ref{reverse2}) and recalling that 
$\Delta m= E_e - E_\nu $,
we get

\begin{equation}
\label{PDB}
\label{reverse3}
\lambda_{e^- +p \rightarrow n + \nu}= \exp{(-\Delta m  / T)}
\lambda_{n+\nu \rightarrow e^- + p}
\end{equation}
which is the expression of the PDB.

When NES is concerned, and correspondingly change from the Fermi-Dirac
distributions to the $n(q)$ ones, the PDB is no longer valid. In this framework, it may be
seen that

\begin{equation}                                    
\exp( -\Delta m/T) (1-n_e(q)) n_\nu(q)=(1-n_\nu(q))n_e(q) +(q-1) \phi,
\end{equation}                                   
where the correction is
                
\bea
\phi =  (1-n_{e,BG}) C_\nu \left( \exp{\left(-\frac{\Delta m}{T}\right)}  
+ \exp{\left(-\frac{E_e}{T}\right) }\right) - \nonumber \\
n_{\nu,BG} C_e \left( \exp{\left(-\frac{\Delta m}{T}\right) 
} + \exp{\left(\frac{E_\nu}{T}\right) } \right)
\eea
and the $C$ factors are as in (\ref{correc-fer}).
In order to get a simplified picture, 
and following the BG case, we further approximate the model
-exactly as was made in the standard case- 
by assuming that during the period of 
freezing, the temperature $T$ is 
low in comparison of the typical energies $E$ 
that contributes in the integrals
for the rates. Hence, we may replace the Fermi-Dirac distributions
by Boltzmann weights (i.e. $n_{\nu , e} \simeq 
\exp{(-E/T)}$) and consistently neglect the
the Pauli blockings (i.e. $(1-n_{\nu ,e}(q)) \simeq 1$). 
Even in the case of nonextensive statistics, Pauli Blockings corrections 
are $1-n_q=1-n_{BG} (1-x^2 \exp{(-x)} (q-1)/2)$, which are neglectable for 
$x \gg 1$ and a first order deviation from $q=1$. In addition, doing this 
allows for the analytical study that follows and which is the objective
of the work. 
In this $x \gg 1$
regime, the one which we are going to use thorough,
we obtain $C_e \simeq C_\nu \exp{(-\Delta m/T)}$, and we recover for the
reverse rates the expression given in (\ref{PDB}) of the PDB.

\subsection{Rates Computation}

Before we finally compute the integrals for the rates we neglect 
the electron mass in
equations (\ref{r1}) and (\ref{r2}) in comparison to the 
energies $E_{\nu ,e}$.
When this is made, the two rates become identical, exactly as they were in BG
statistics. Then we obtain:
                       
\begin{equation}                 
\lambda_{\nu+n \rightarrow p+e^-}=
\lambda_{\nu+n \rightarrow p+e^-}^{standard}+
\left( 480 T^5 +2 \times 84 T^4 Q + 18 T^3 Q^2\right)\,(1-q) A 
\end{equation}
where,

\begin{equation}                 
\lambda_{\nu+n \rightarrow p+e^-}^{standard}=
\left(4! T^2 + 2 \times 3!T Q +2! Q^2\right) \, A T^3.
\end{equation}                           
Concerning the rate of free decay of neutrons, we note that 
NES does not present any
change within the context of previous approximations. 
That is due to the disappearence of all
distributions functions from (\ref{r3}). In that equation we cannot neglect the 
mass of the electron and the standard result holds,

\begin{equation}                 
\frac{1}{\tau} = \lambda_{n \rightarrow p+e^- + \hat \nu} = 0.0157 A 
\Delta m^5.
\end{equation}                           
This enable to eliminate $A$ in favor of the measured quantity $\tau$,

\begin{equation}                 
A= \frac{a}{\tau} \frac 14 \Delta m^5,\;\;\;\;\;\;\;a=255.
\end{equation}                           
All along in this section, we shall 
neglect the free neutron decay rate when computing
the total 
$\lambda_{np}$. Thus, we have

\begin{equation}                 
\lambda_{np}=2\lambda_{\nu+n \rightarrow p+e^-}.
\end{equation}
Finally, by using the variable $y=\Delta m /T$, we get

\begin{equation}                 
\lambda_{np}=\frac{a}{\tau y^5} \left [
\left( 12 +6 y+y^2 \right) + (1-q) \left( 240 +84 y +9 y^2 \right) 
\right],
\end{equation}
for the neutron-proton rate and

\begin{equation}                 
\lambda_{pn}=\exp{(-y)} \lambda_{np},
\end{equation}                      
for the reverse reaction.

\subsection{Evolution of the neutron abundance}

In equation (\ref{X4}), we change now variables to the scaled temperature  $y$
and obtain,

\begin{equation}
\label{X4y}
X(y)= \frac {\lambda_{pn}(y)}{\Lambda(y)}-
\int_{0}^y  dy^{\prime} I(y,y^\prime) \frac{d}{dy^{\prime}} \left( 
\frac{\lambda_{pn}(y^{\prime})}{ \Lambda(y^{\prime})}\right).
\end{equation}                  
The integrating factor now becomes,

\begin{equation}
I(y,y^\prime) =\exp {\left( - \int_{y^\prime}^y    d\hat y \left( \frac
{d \hat t}{d \hat y} \right) \Lambda(\hat y) \right)}.
\end{equation}                                                    
To evaluate the jacobian ${d \hat t}/{d \hat y}$, we need to recall that the scale
factor of the universe, $R$, in a Friedmann-Robertson-Walker metric, behaves
as $R \simeq 1/T$, independently of the statistics \cite{Barraco}. Therefore,
$\dot T /T = - \dot R /R$, and the right hand side is given by Einstein 
equations:

\begin{equation}
\frac {\dot R}{R} = \left( \frac{ 8 \pi G}{3} \rho \right)^{\frac 12}.
\end{equation}
Here $\rho$ is the energy density for relativistic particles in NES. This 
may be easily computed, if the distributions functions are known, by

\begin{equation}
\rho = \frac{g}{2 \pi^2} \int_0^{\infty} E^3 n(q) dE,
\end{equation}
where $g$ is the degeneracy factor. The result is 

\begin{equation}
\label{rho}
\rho=
\rho_{bosons} +\rho_{fermions} = \frac{\pi^2}{30} g T^4 + \frac{1}{2\pi^2}
\left(  40.02 \, g_b+ 34.70 \, g_f\right)\,T^4 (q-1) ,
\end{equation}
where $g=g_b+7/8 g_f$. 
At high enough temperatures the energy density of the universe is essentially
dominated by $e^-$, $e^+$, $\nu$ and $\gamma$'s. Interactions among these
particles keep all them nearly the same temperature. Accordingly, we set,                         
$g_b=2$, $g_f=2+2+2 \times 3=10$ and $g=43/4$. With these values we get,                       
                                                    
\begin{equation}
\label{rho2}
\rho=\frac{\pi^2}{30} g T^4 + \frac{1}{2\pi^2} 21.63 \, T^4\, (q-1) ,
\end{equation}                                                    

With this in mind, and using                                                     
$\dot T /T = - \dot a /a$, we obtain

\begin{equation}
\frac{dT}{dt} = - \left[ \left( \frac{4\pi^3 G g}{45} \right) +
\left( \frac{ 4 G}{3 \pi} \right) 21.63 (q-1) \right]^{\frac 12} T^3
\end{equation}                                                      
which yields, to first order in $(q-1)$ to,

\begin{equation}
\frac{dt}{dy} = \frac {y}{\Delta m^2 }\left( \frac{45}{4\pi^3 G g} \right)^{\frac 12}
 \left[ 1- \frac 12 \left( \frac{ 45 }{3 \pi^4} \right) \frac{21.63}{g} (q-1) 
 \right].
\end{equation}                                                      
We shall call the constants,

\begin{equation}
b=\left( \frac{45}{4\pi^3 G g} \right)^{\frac 12} \frac{1}{\tau \Delta m^2} a
\end{equation}                                                      
and 

\begin{equation}
c= b  \left( \frac{ 45 }{6 \pi^4} \right) \frac{21.63}{g}.
\end{equation}                                                      
With all above, the integrating factor becomes,

\begin{equation}
I(y,y^\prime) =\exp {(K(y)-K(y^\prime))},
\end{equation}                                 
where,

\begin{eqnarray}
K(y) =-b \int dy   \left[ \left(  \frac {12}{y^4} +
\frac {6}{y^3}+ \frac {1}{y^2} \right) \left( 1 + c(1-q) \right)
+ 
 \left(  \frac{240}{y^4} + \frac {84}{y^3}+ \frac {9}{y^2} \right) (1-q) 
 \right] \times \nonumber \\
 (1+\exp{(-y)}).
\end{eqnarray}                                 
This integrates to give,

\begin{eqnarray}
\label{KK}
K(y) = b  \left(  \frac {4}{y^3} + \frac {3}{y^2}+ \frac {1}{y} +
\left(  \frac {4}{y^3} + \frac {1}{y^2} \right) \exp{(-y)} \right)
 \left( 1 + c(1-q) \right) \nonumber \\
+b\left(  \frac {80}{y^3} +
\frac {42}{y^2}+ \frac {9}{y} +
\left(  \frac {80}{y^3} +
\frac {2}{y^2}+ \frac {7}{y} \right) \exp{(-y)} - 7 Ei(1,y) \right) 
(1-q) .
\end{eqnarray}          
where $Ei$ stands for the exponential integral 
\footnote{ The exponential integral is defined by $Ei(n,x) = \int_1^\infty
\frac{\exp{(-xt)}}{t^n} dt $ for $Re(x) >0$.}.
Introducing,

\begin{equation}
X_{eq}=\frac{\lambda_{pn}(y)}{\Lambda(y)}=\frac{1}{1+\exp{(y)}},
\end{equation}
the neutron abundance ratio reads:

\begin{equation}
X(y)=X_{eq}+ \int_o^y dy^{\prime} \exp{(y^{\prime})} X_{eq}(y^{\prime})^2
\exp{[K(y)-K(y^{\prime})]}.
\end{equation}
The exact previous result may be expanded again 
to first order in $(q-1)$. This gives,

\bea
X(y)=X_{standard}    + 
(1-q) \exp{(bA(y))} 
(bB(y)+bcA(y)) 
\times \nonumber \\
\int_o^y dy^{\prime} \exp{(y^{\prime})} X_{eq}(y^{\prime})^2
\exp{(bA(y^\prime)}) \nonumber \\
-(1-q) \exp{(bA(y))}     
\int_o^y dy^{\prime} \exp{(y^{\prime})} X_{eq}(y^{\prime})^2
\exp{(bA(y^\prime)}  (bB(y^\prime)+bcA(y^\prime)) ,
\eea
where,

\begin{equation}
X_{standard} = X_{eq}(y) + \exp{(bA(y))} 
\int_o^y dy^{\prime} \exp{(y^{\prime})} X_{eq}(y^{\prime})^2
\exp{(bA(y^\prime))},
\end{equation}
and the functions $A$ and $B$ are:

\begin{equation}
A(y)= \left(  \frac {4}{y^3} + \frac {3}{y^2}+ \frac {1}{y} +
\left(  \frac {4}{y^3} + \frac {1}{y^2} \right) \exp{(-y)} \right)
\end{equation}
\begin{equation}
B(y)= \left(  \frac {80}{y^3} +
\frac {42}{y^2}+ \frac {9}{y} +
\left(  \frac {80}{y^3} +
\frac {2}{y^2}+ \frac {7}{y} \right) \exp{(-y)} - 7 Ei(1,y) \right) .
\end{equation}
We can numerically compute the 
integrals that appears here. Using explicit values 
for the constants                                                                     
$b=0.252$ and $c=0.00244$ with the mean life                                                                      
of the neutron given by $\tau=889.8s$ we get                                                                     
the curves of Fig. 1. There, it is shown the standard value of the
neutron to baryon ratio and the correction due to NES. In Fig. 2.
are shown different explicit computations for a range of the $q$ parameter.
Finally, it may be seen that  $X(y)$ asymptotes to

\begin{equation}
X(y=\infty)=0.15+(q-1) 1.15,
\end{equation}
for very low temperatures.

\subsection{Possible dependence on the number of neutrino types}

We briefly assess here the possible dependence on the number of neutrino 
types. To do this we note that such a variation would introduce changes in the
coefficients $b$ and $c$ quoted above.  For three neutrino families, $g=43/4$
and the addition of one more family change it in an amount $\delta g=7/4$.
Then, the standard case  coefficient $b$ is affected as,

\begin{equation}
\frac {\delta b}{b}= -\frac 12 \frac{\delta g}{g}
\end{equation}
and the non-extensive related as,

\begin{equation}
\frac {\delta c}{c} = -\frac 32 \frac{\delta g}{g}.
\end{equation}
Introducing these new values in (\ref{KK}) and making all numerical
computations again, it is possible to obtain the corrections
due to a different number of neutrino families. In the sake of conciness,
we shall explicitly skip these computations. Instead we note that, while
in the standard case, the addition of one more neutrino family would involve
an increment of the energy density of the universe and a speed up in its
expansion, in the general case, the correction will depend on the sign of the
factor $(q-1)$.

\section{Towards computing Neutron Decay Corrections} 

We have already solved the evolution of the neutron 
abundance neglecting the neutron
decay. As in the 
standard formalism, it is useful to change notation by using an
overbar for the result just obtained, $X(y) \rightarrow \bar X(y)$.
Including the effects of the free neutron                                              
decay in the rate equation we would get,
 
\begin{equation}                               
\label{ult}
 X(t)=\exp{(-t/\tau)} \bar X(t),
 \end{equation}
because $\bar X$ does not vary much during the period in which neutrons
decay.                                              
In the capture time $t=t_c$, 
when the temperature drops somewhat below the deuteron
binding energy ($\epsilon_D=2.62 {\rm MeV}$), 
the neutrons are captured in deuterons.
Then, deuterons collide and almost all neutrons present at $t=t_c$
are converted into $^4\!He$. Inserting this time                                              
in (\ref{ult}) and using the asymptotic value of $\bar X(y)$ just derived would
yields, thus,  half  of the mass fraction of helium produced in the early universe.
In order  to obtain a precise value of $t=t_c$ or $t_{c,q}$ for the 
nonextensive result, we have to analize the set
of the basic reactions:

\beq
\label{reac4}
n + p \Longleftrightarrow    D + \gamma
\eeq
\beq
\label{reac5}
D+ D \Longleftrightarrow   p + T
\eeq
\beq               
\label{reac6}
T+ D  \Longleftrightarrow  ^4\!\!He + n.
\eeq                                    
Doing this in the standard case \cite{Bernstein} yields  a value of $t_c$
such as the correction $\exp{(-t_c/\tau)} \simeq 0.8$.
Although the analysis of (\ref{reac4},\ref{reac5},\ref{reac6})
in a NES framework 
is beyond the scope of the present work, we shall establish here the
basis on which this could be done in the future.
For this, we shall study the process of annihilation of pairs $e^- , e^+$.

\subsection{Conserved comoving quantities}

We know that, when the temperature of the universe is low compared with
the electron mass, electrons and positrons annihilate heating the photons 
and, morover, changing the effective quantity  of degrees of freedom.
Then, one of 
the assumptions of the previous section, i.e. $T=T_e=T_\nu=T_\gamma$,
is no longer valid and $T_\gamma$ will be different from the decoupled neutrino
temperature. We shall need the connection between these temperatures. To
compute it, we begin by noting that, at these times, the number of neutrinos
in a comoving volume is fixed, independently of the statistics. Effectively,
after the moment in which the collision rate for neutrinos is lower than the
expansion of the universe, $\dot R/R$, they cannot be 
neither created nor destroyed.
As the number of neutrinos is given by,

\beq
n_\nu (q)= \frac{g}{2 \pi^2} \int_0^\infty E^2 <\hat n>_q =
\frac {3}{4 \pi^2} \zeta (3) g_\nu T^3 + 
(q-1) \frac {g_\nu}{2 \pi^2} 5.50 T^3
\eeq
the relationship,

\beq
n_\nu(t) R(t)^3 = const.,
\eeq
or, equivalently,

\beq                                
\label{law}
T_\nu(t) R(t) = const.,
\eeq
continues holding.

>From the full set of Einstein field equations, it may be established that

\beq
d(\rho(q) R^3) = -p(q) d(R^3),
\eeq
this may also be written as,

\beq
\label{rel1}
\frac{d}{dT} \left[\left( \rho(q) + p(q) \right)R^3\right] = (R^3) \frac{dp(q)}{dT}.
\eeq                                                                              
But we may compute $p(q)$, as was done with the energy density,
obtaining $p(q)=\rho(q)/3$ for the 
relativistic gas. Using this and (\ref{rel1}),
we get 

\beq
\label{rel2}
\frac{d}{dT} \left[ \frac{\rho(q) + p(q) }{T} R^3\right] = 0,
\eeq                                                                            
as can be cheked by direct differentiation.  
This means, that in general case, $(\rho(q) + p(q))/ T$
is a conserved quantity in a comoving volume.
When $q=1$ the previous expression stands for the entropy of the system.

\subsection{Pair annihilation}

We may now consider that pair annihilation occurs at a temperature $T$,
and then indicate with symbols $-$ and $+$,
quantities before and after $T$. From all above, we have

\beq
\label{rel3}
\frac{\rho(q)_- + p(q) _-}{T_-}=\frac{\rho(q) _+ + p(q)_+ }{T_+}.
\eeq
Here $\rho(q)_{-,+}$ is as in (\ref{rho}) with the corresponding values of 
$g_{-,+}$ and $g_{b,f,-,+}$. Due to the removal of pairs, we have that 
$g_+ < g_-$. We may now expand (\ref{rel3}), to first order in $(q-1)$,
obtaning the relationship,

\beq
T_+^3 = T_-^3 \left( \frac {g_-}{g_+} \right) \left[ 1 + (q-1) \delta \right],
\eeq
where, 

\beq
\delta = (40.02 g_{b,-} + 34.70 g_{f,-} ) \frac{30}{2\pi^4 g_-} - 
(40.02 g_{b,+} + 34.70 g_{f,+} ) \frac{30}{2\pi^4 g_+}.
\eeq
Until the moment of $e^-, e^+$ annihilation, the gas composed by
$e^-, e^+$and $\gamma$ follows a law of identical form that 
(\ref{law}), as such temperatures, being $T_e=T_\gamma$. But
after the annihilation, where only $\gamma 's$ remain, we have

\beq
\label{rel4}
T_\gamma = T_\nu \left( \frac {11}{4} \right) ^\frac 13 \left[ 1+ (1-q) 0.013 
\right],
\eeq    
where the corresponding values $g_-=g(e^-,e^+,\gamma)=11/2$
and $g_+=g(\gamma)=2$ were used.

\subsection{Neutron capture time}

To finally compute the capture time $t_c$, we use that

\beq
\frac{1}{T_\nu} \frac{dT_\nu}{dt}= - \left( \frac {8 \pi G}{3} \rho(q) \right)^
\frac 12  ,
\eeq
from where

\beq          
\label{t}
t= \int_{T_\nu}^\infty      \frac{dT_\nu^\prime}{T_\nu^\prime}
 \left( \frac {3}{8\pi G \rho(q)} \right)^
\frac 12 
\eeq
follows. Here $\rho(q)$ adquires its                                              
low energy form

\beq
\rho_0=\left[ g_\nu \frac{\pi^2}{30} T_\nu^4 +(q-1) 34.70 T_\nu^4
\frac{g_\nu}{2\pi^2}
\right]
+\left[ g_\gamma \frac{\pi^2}{30} T_\gamma^4 +(q-1) 40.02 T_\gamma^4
\frac{g_\gamma}{2\pi^2}
 \right]  ,
\eeq                                              
where $g_\nu =3$ and $g_\gamma=2$. Recalling the relationship (\ref{rel4})
we get, to first order in $(q-1)$,

\beq
\rho_0=g_{eff} \frac{\pi^2}{30} T_\nu^4 +(q-1)  T_\nu^4 14.84
\eeq
where $g_{eff}=g_\nu + g_\gamma (11/4)^{4/3}$.

In the neighbourhood of the time of the neutron capture , an approximate
evaluation of (\ref{t}) is in order. Using the expression for
$\rho_0$ just derived, we get

\beq
t=\left( \frac{ 45}{16 \pi^3 g_{eff} G} \right)^\frac 12 \frac{1}{T_\nu^2}
(1+(1-q) 1.74) +t_0,
\eeq                
where $t_0$ is an additional integration constant and the corrective
number 1.74 arise from the first order evaluation of $\rho(q)^\frac 12 $.
Recalling then the relationship between the temperatures, we may express

\beq                                                                      
\label{t2}
t=\left( \frac{ 45}{16 \pi^3 g_{eff} G} \right)^\frac 12 \frac{1}{T_\gamma^2}
(1+(1-q) 1.76) +t_0,
\eeq
We may neglect $t_0$ in a first approximation or obtain it
from a perturbation analysis of (\ref{t}), \cite{Bernstein}.

Now, a comprehensive analysis of the reactions 
(\ref{reac4},\ref{reac5},\ref{reac6}) would yield a value for $T_\gamma$, 
for which almost all
nucleons are converted into helium . Using then (\ref{t2}), the evaluation
of $t_c$ follows and  thus, the mass fraction of primordial helium.
In passing through NES, we have to note that a fundamental tool 
in this analysis,
i.e. Saha formulae, is modified in a non-trivial way. This entails
modifications in the abundance fractions
that may in turn change their ratios, which are the
magnitudes on which the set depends. So, although phenomenological fits
for the rates may be considered, we acknowledge here that further research
must be done
to analytically solve for the reactions, 
in the case this be ultimately possible.
This analysis is far out from the present work and we shall not discuss this
problem in deeper detail.

\section{Conclusions and comparison with others related results}

We have presented a comprehensive study of the primordial 
neutron-baryon ratio in the context of 
non-extensive statistics.
This work, thus, extends previous advances on the helium nucleosynthesis
production problem \cite{TORRES_T1} and is the result of a detailed analysis
of the weak interaction rate and the energy densities corrections
at first order in the non-extensive parameter $(q-1)$.  We have also studied
some important cosmological scenarios such as the possible
existence of conserved comoving quantities, temperatures relationships and
pair annihilation, together with pointing out where further research is needed.
These other results will be used elsewhere to put new stringent bounds
on the non-extensivity of the universe by using high precision satellite
cosmological measurements of the cosmic microwave background and
new galaxy surveys.

Even without getting a precise analytical value for the Helium synthesis,
it is possible to obtain a first insight in the problem by 
considering it as exactly twice of the neutron-baryon asymptotic value,
weighted by the free neutron decay:

\beq
Y_p^{theoretical} \sim 2 X(t)
\eeq
where $X(t)$ is given by equation (56).
But as mentioned above, in order to put an exact bound 
we would need the value of $t_{c,q}$. However, even considering that 
$T_\gamma$ is the same as in the standard formalism
($T_\gamma \simeq 0.086{\rm MeV}$) we could use (\ref{t2}) to get the
correction in $t_c$. Using afterwards the first order expansion of (\ref{ult})
(which contains equation (53))
we would get the stringent bound: 

\beq                 
\label{BOU}
|1-q| < 2.6 \times 10^{-4}. 
\eeq
In order to get this bound, 
we have imposed that the non-extensive corrections (coming from equations
(75) and (53) and the theoretical value of $Y_p$ as explained above) 
be less than
the observational error in the
primordial Helium production, which is doubled so as to
neglect the difference
between the central 
values of the analytical and observational magnitudes \cite{Steigman}, i.e.:

\beq
Y_p^{observational}=0.23 \pm 0.02.
\eeq
This abundance is a very safe estimate of the primordial Helium production,
which in fact is a very complex observable  magnitude, 
see for instance the recent review by Schramm and Turner \cite{ST} 
for details. Besides, 
it encodes
other possible uncertainties such as the value of the cosmological constant
or the exact number of neutrino species. 

The bound (\ref{BOU}) is
comparable with that already obtained in Ref. \cite{TORRES_T1}.
Although it should be recalled that such derivation was different from this
and the method followed skip the uncertainties of the neutron capture time
here analyzed.  In addition, the weak interaction rates were only approximately 
treated in \cite{TORRES_T1} while here they were studied in full detail.

In a recent letter \cite{ALEMANY}, it was suggested that the universe as a whole
could behave with a non-extensive parameter roughly equal to 0.54. Such a value seems to be 
too high as to resist early universe tests as the one presented here.

Therefore, this bound, obtained in the simplest situation so as to admit
an analytical solution, pertains to those imposed in the context 
of early cosmological models, 
and from this point of view, enlarge the problems that
non-extensivity would have as a correct description
of the physical universe thorough all eras of cosmic evolution.
Very recently, other works have also been devoted to study possible deviation
from the Boltzmann-Gibbs statistics in cosmological and other situations. 
Among them, and apart from Ref.
\cite{TORRES_T1} already commented, we should note the following ones:

\begin{itemize}

\item Plastino, Plastino and Vucetich \cite{P2V} studied the value of the
Stefan-Boltzmann constant in the new framework. A comparison with the
experimental value leads to $|q-1| \leq 0.67 \times 10^{-4}$. They also used
data from the FIRAS experiment on the COBE satellite concening the microwave
background radiation, from there they obtain $|q-1| \leq 5.3 \times 10^{-4}$.

\item Tsallis, Sa Barreto and Loh \cite{Tsallis3} also generalized the
Planck radiation law and used satellite data from COBE to constrain the
non-extensive parameter. They found $|q-1| < 3.6 \times 10^{-5}$.

\item Tirnakli, B\"uy\"ukkili\c{c} and Demirhan \cite{TIRNAKLI,TIR2} 
reworked the generalized quantal distributions 
functions and also obtain a similar bound: $|q-1| \leq 0.41 \times 10^{-4}$.

\end{itemize}

These works, among others,  suggest that non-extensivity is highly
constrained concerning the physics of the early universe.

\subsection*{Acknowledgments}
The authors acknowledge partial support from CONICET and valuable
discussions with A. Plastino and U. Tirnakli during the course 
of this research. Useful insights from an anonymous referee helped to
improve the presentation and are gratefully acknowledged.
D.F.T. was supported by a Chevening Scholarship of the British Council.

\newpage


\begin{figure}  
\label{fig.1}
\caption{Behaviour of the primordial neutron to baryon 
ratio in standard Boltzmann-Gibbs
statistics (curve (1)) and correction due to non-extensivity
weighted by $(q-1)$, (curve (2)). 
The complete behaviour of the neutron-baryon ratio in NES
is obtained making the sum of curve (1) plus $(q-1) \times $  curve (2).}
\begin{center}
    \leavevmode
   \epsfxsize = 10cm
    \epsfysize = 8cm
    \epsffile{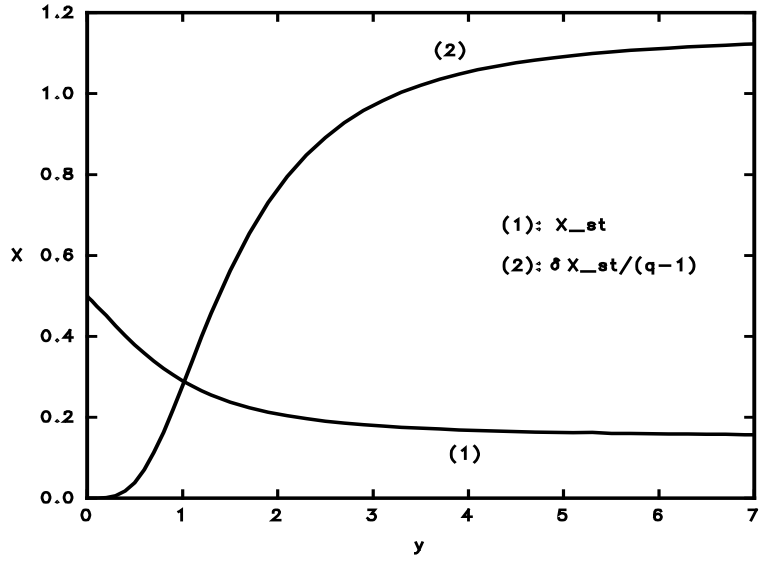}
    \end{center}

\end{figure}

\begin{figure}  
\label{fig.2}
\caption{
Neutron to baryon 
ratio relationship for different non-extensive parameter
$q$.}
\begin{center}
    \leavevmode
   \epsfxsize = 10cm
     \epsfysize = 8cm
    \epsffile{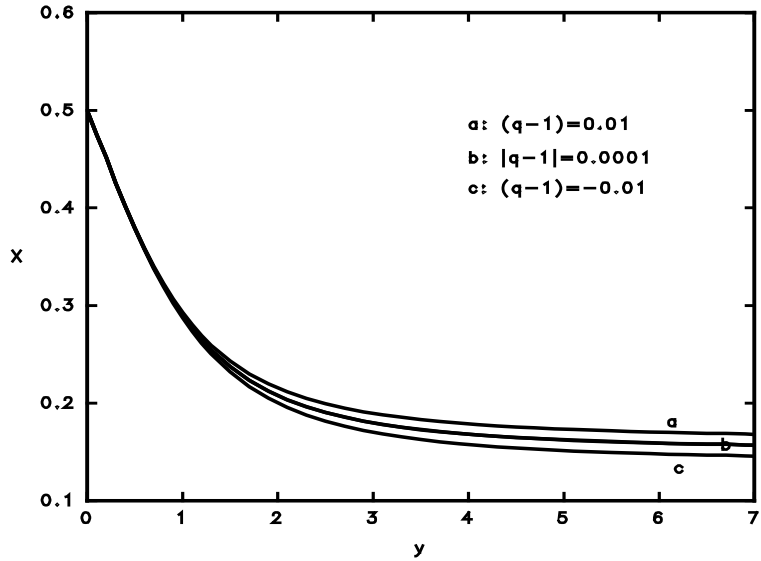}
    \end{center}

\end{figure}


\begin{references}

\bibitem{t2} C. Tsallis, J. Stat. Phys. {\bf 52}, 479 (1988).

\bibitem{t1} C. Tsallis, Fractals {\bf 6}, 539 (1995).

\bibitem{PHYSA} C. Tsallis, Physica {\bf A221}, 227 (1995).

\bibitem{WEB} http://tsallis.cat.cbpf.br

\bibitem{t3} E. M. F. Curado and C. Tsallis, J. Phys. {\bf A24}, L69 (1991);
Corrigenda: {\bf A24}, 3187 (1991)  and {\bf A25}, 1019 (1992).

\bibitem{PLAS-ENTRO} A. Plastino and A. R. Plastino, Phys. Lett. {\bf A226},
257 (1997).

\bibitem{t4} A. R. Plastino and A. Plastino, Phys. Lett. {\bf A177},
177 (1993).

\bibitem{t5} A. R. Plastino and A. Plastino,
Phys. Lett. {\bf A174}, 384 (1993).

\bibitem{t6} A. R. Plastino and A. Plastino, Phys. Lett. {\bf A193},
251 (1994).

\bibitem{t7} P. A. Alemany and D. H. Zanette, Phys. Rev. {\bf E49},
956 (1994).

\bibitem{BOGHO} B. M. Boghosian, Phys. Rev. {\bf E53}, 4754 (1996).
\bibitem{solar} G. Kaniadakis, A. Lavagno and P. Quarati,
Phys. Lett. {\bf B369}, 308 (1996).

\bibitem{GRIGERA} J. R. Grigera, Phys. Lett. {\bf A217}, 47 (1996).

\bibitem{KIM} P. Jund, S. G. Kim and C. Tsallis, Phys. Rev. {\bf B52}, 50 (1995).

\bibitem{SAMPAIO} L. C. Sampaio, M. P. Alburquerque and S. Menezes, 
Phys. Rev. {\bf B55}, 5611 (1997).

\bibitem{Tsallis3} C. Tsallis, 
F. Sa Barreto and E. D. Loh, Phys. Rev. {\bf E52}, 1447 (1995).

\bibitem{P2V} A. R. Plastino, A. Plastino and H. Vucetich, Phys. Lett.
{\bf A207}, 42 (1995).

\bibitem{Barraco} V. H. Hamity and D. E. Barraco, Phys. Rev. Lett. {\bf 76},
25 (1996).

\bibitem{TORRES_T1} D. F. Torres, H. Vucetich and A. Plastino,
Phys. Rev, Lett. {\bf 79}, 1588 (1997); E{\bf 80}, 3889 (1998).

\bibitem{Bernstein} J. Bernstein, L. S. Brown and G. Feimberg, 
Rev. Mod. Phys. {\bf 61}, 25 (1989).

\bibitem{Casas2} J. A. Casas, J. Garcia-Bellido and M. Quir\'os, Phys.
Lett. {\bf B278}, 94 (1992). 

\bibitem{TORRES_N} D. F. Torres, Phys. Lett. {\bf B359}, 249 (1995).

\bibitem{TORRES_W} L. A. Anchordoqui, D. F. Torres and H. Vucetich,
Phys. Lett. {\bf A222}, 43 (1996).

\bibitem{Steigman}  G. Steigman  Nuc. Phys. (Proc. Suppl.) {\bf B37c}, 68 
(1995).

\bibitem{Kolb} E. W. Kolb, M. S. Turner, {\it The Early Universe}
(Addison-Wesley Publishing Company, 1990).

\bibitem{Lucchin} P. Coles and 
F. Lucchin, {\it Cosmology}, (John Wiley \& Sons, 1995).

\bibitem{Weimberg} S. Weimberg, {\it Gravitation and Cosmology}
(John Wiley, New York, 1972).

\bibitem{ALEMANY} P. A. Alemany, Phys. Lett. {bf A235}, 452 (1997).

\bibitem{TIRNAKLI} U. Tirnakli, F. B\"uy\"ukkili\c{c} and D. Demirhan,
{\it Quantal Distribution Functions and Some Alternative Bounds upon
Nonextensive Parameter}, to appear en Phys. Lett. {\bf A}.

\bibitem{ST} D. N. Schramm and M. S. Turner, 
Rev. Mod. Phys. {\bf 70}, 303 (1998).  
                                                     
                                                       
\bibitem{TIR2} U. Tirnakli, F. B\"uy\"ukkili\c{c} and D. Demirhan,
Physica {\bf A240}, 657 (1997).





\end{references}
\end{document}